\documentclass[letterpaper]{article} 
\usepackage{aaai2026}  
\usepackage{times}  
\usepackage{helvet}  
\usepackage{courier}  
\usepackage[hyphens]{url}  
\usepackage{graphicx} 
\usepackage{natbib}  
\usepackage{caption} 
\usepackage{algorithm}
\usepackage{algorithmic}

\usepackage{newfloat}
\usepackage{listings}
\DeclareCaptionStyle{ruled}{labelfont=normalfont,labelsep=colon,strut=off} 
\floatstyle{ruled}
\newfloat{listing}{tb}{lst}{}
\floatname{listing}{Listing}
\usepackage{graphicx}
\usepackage{booktabs}
\usepackage{array}    
\usepackage{pifont}
\newcommand{\cmark}{\ding{51}}  
\usepackage{bm}
\newcommand{\cmarkdark}{\bm{\cmark}}
\newcommand{\xmark}{\ding{55}}  
\usepackage{makecell}
\usepackage{caption}

\usepackage{xcolor}
\definecolor{customblue}{RGB}{0, 0, 255}

\usepackage{tabularx}
\usepackage{multirow} 
\usepackage{subcaption}
\usepackage[most]{tcolorbox}

\usepackage{array}
\newcolumntype{L}[1]{>{\raggedright\arraybackslash}p{#1}}
\usepackage{placeins}

\title{MASH: A Multiplatform and Multimodal Annotated Dataset for \\
Societal Impact of Hurricane}

\author{
    Ruichen Yao\textsuperscript{\rm 1}, 
    Aslanbek Murzakhmetov\textsuperscript{\rm 1}\textsuperscript{\rm 2}\equalcontrib, 
    Raaghav Pillai\textsuperscript{\rm 1}\equalcontrib, 
    Aliya Maussymbayeva\textsuperscript{\rm 1}\textsuperscript{\rm 5}\equalcontrib, 
    Zelin Li\textsuperscript{\rm 1},\\
    Yifan Liu\textsuperscript{\rm 1}, 
    Yaokun Liu\textsuperscript{\rm 1}, 
    Lanyu Shang\textsuperscript{\rm 3}, 
    Yang Zhang\textsuperscript{\rm 4}, 
    Na Wei\textsuperscript{\rm 1}, 
    Ximing Cai\textsuperscript{\rm 1}, 
    Dong Wang\textsuperscript{\rm 1} 
}
\affiliations{
    \textsuperscript{\rm 1}University of Illinois Urbana-Champaign, Champaign, Illinois, USA\\

    \textsuperscript{\rm 2}M.Kh. Dulaty Taraz University, Taraz, Jambyl Region, Kazakhstan\\
    \textsuperscript{\rm 3}Loyola Marymount University, Los Angeles, California, USA\\
    \textsuperscript{\rm 4}Miami University, Oxford, Ohio, USA\\
    \textsuperscript{\rm 1}\{ryao8, aslanbek, pillai7, aliyam, zelin3, yifan40, yaokunl2, nawei2, xmcai, dwang24\}@illinois.edu\\
    \textsuperscript{\rm 2}an.murzakhmetov@dulaty.kz, \textsuperscript{\rm 3}lanyu.shang@lmu.edu, \textsuperscript{\rm 4}zhang981@miamioh.edu, \textsuperscript{\rm 5}maussymbayevaaliya@gmail.com\\
}

\usepackage{bibentry}

\begin{document}

\maketitle
\begin{abstract}
        Natural disasters cause multidimensional threats to human societies, with hurricanes exemplifying one of the most disruptive events that not only caused severe physical damage but also sparked widespread discussion on social media platforms. Existing datasets for studying societal impacts of hurricanes often focus on outdated hurricanes and are limited to a single social media platform, failing to capture the broader societal impact in today's diverse social media environment. Moreover, existing datasets annotate visual and textual content of the post separately, failing to account for the multimodal nature of social media posts. To address these gaps, we present a multiplatform and \textbf{M}ultimodal \textbf{A}nnotated Dataset for \textbf{S}ocietal Impact of \textbf{H}urricane (\textbf{MASH}) that includes 59,607 relevant social media data posts from Reddit, TikTok, and YouTube. In addition, all relevant social media data posts are annotated in a multimodal approach that considers both textual and visual content on three dimensions: Humanitarian Classes, Bias Classes, and Information Integrity Classes. To our best knowledge, MASH is the first large-scale, multi-platform, multimodal, and multi-dimensionally annotated dataset centered on hurricane disasters. In addition, we introduce an online platform that supports interactive data exploration, provides preliminary analytical results, and allows users to share their insights regarding the societal impacts of hurricanes. We envision that MASH can contribute to the study of hurricanes' impact on society, such as disaster response, disaster severity classification, public sentiment analysis, disaster policy making, and bias identification. 
        The dataset is publicly available
        under the Creative Commons Attribution 4.0 (CC BY 4.0) license. 
        \begin{links}
    \link{Dataset}{https://huggingface.co/datasets/YRC10/MASH}
  \link{Online Platform}{https://hurricane.web.illinois.edu}

\end{links}


\end{abstract}
\section{Introduction} \label{Introduction}

\begin{table*}[t]
\centering
\footnotesize
\setlength{\tabcolsep}{2pt}
\renewcommand{\arraystretch}{2}

\begin{tabular}{c|c|c|c|c|c|c}
\toprule[1.5pt]
\textbf{Dataset} & \textbf{\makecell{Collection \\ Time}} & \textbf{\makecell{Hurricanes \\ Anno Amount}} & \textbf{\makecell{MultiModal}} & \textbf{\makecell{Humanitarian \\ Anno}} & \textbf{\makecell{Bias \\ Anno}} & \textbf{\makecell{Info Integ \\ Anno}}  \\ \hline
\makecell{Sandy 2012~\cite{sandy2012}} & \makecell{2012} & 2,000 & \xmark  & \cmark  & \xmark & \xmark  \\ \hline
\makecell{CrisisMMD~\cite{crisismmd}} & \makecell{2017} & 12,041 & \cmark  & \cmark & \xmark & \xmark  \\ \hline
\makecell{HIM-Twitter~\cite{HIM-Twitter}} & \makecell{2017} & 50,217 & \cmark  & \cmark & \xmark & \xmark  \\ \hline
\makecell{TweetDIS~\cite{TweetDIS}} & \makecell{2012 - 2017} & 5,962 & \xmark  & \cmark & \xmark & \xmark  \\ \hline
\makecell{Eyewitness Messages~\cite{eyewitness}} & \makecell{2016 - 2018} & 2,000 & \xmark  & \cmark & \xmark & \xmark  \\ \hline
\makecell{MEDIC~\cite{MEDIC}} & \makecell{2017 - 2018} & 6,686 & \xmark  & \cmark & \xmark & \xmark \\ \hline
\makecell{Natural Hazards~\cite{NaturalHazards}} & \makecell{2012 - 2019} & 7,823 & \xmark  & \xmark & \xmark & \xmark  \\ \hline
\makecell{HumAID~\cite{humaid}} & \makecell{2016 - 2019} & 45,581 & \xmark  & \cmark & \xmark & \xmark  \\ \hline 
\textbf{\makecell{MASH (Ours)}} & \textbf{\makecell{2024 (latest)}} & \textbf{59,607}& \textbf{\cmarkdark}  & \textbf{\cmarkdark} & \textbf{\cmarkdark} & \textbf{\cmarkdark}  \\ 
\bottomrule[1.5pt]
\end{tabular}
\caption{Comparison with Existing Hurricane Disaster Datasets}

\label{tab:comparison}
\end{table*}


\begin{figure}[t]
    \centering
    \begin{minipage}{0.5\textwidth} 
        \centering
        \includegraphics[width=0.9\textwidth]{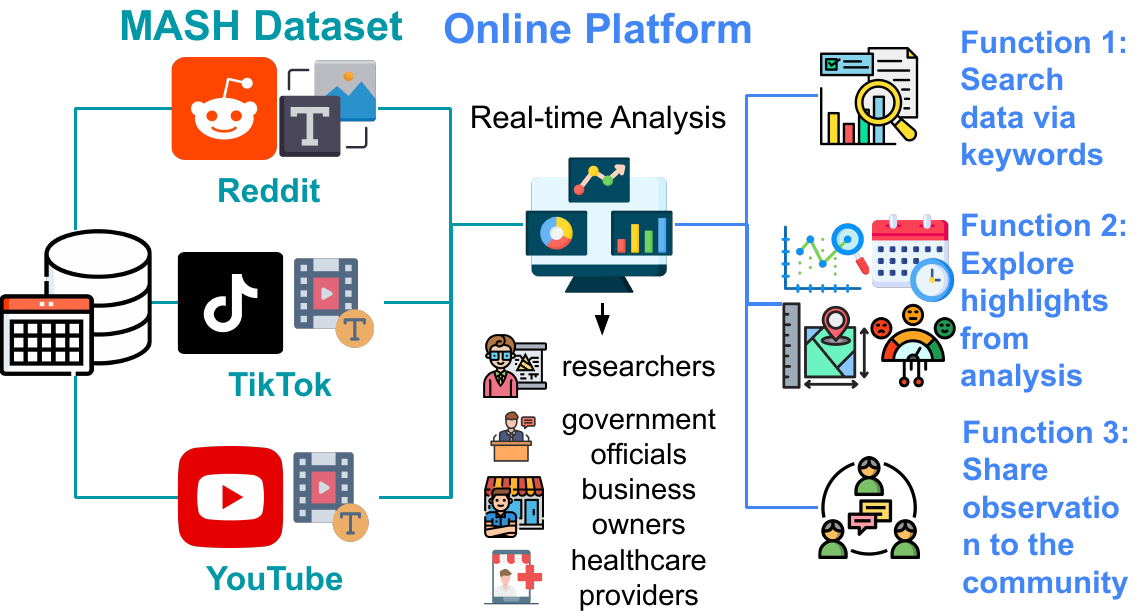}
        \caption{MASH Dataset and Online Platform} 
        \label{fig:overview} 
    \end{minipage}
\end{figure}

Natural disasters have been a persistent threat to human societies, with hurricanes standing out as one of the most representative disasters that often cause severe destruction and significant societal impacts~\cite{shang2024socialdrought, shang2025side, crisismmd,humaid}.
For example, Category 4 Hurricane Helene and Category 5 Hurricane Milton struck the United States in 2024, resulting in at least 250 casualties and over \$300 billion in economic losses~\cite{2024_Hurricane_Season, ScienceHurricane}. More importantly, beyond their physical destruction, these hurricanes triggered extensive social media interactions that shaped public understanding, trust in official information, and coordination of community aid, especially among socially vulnerable groups such as low-income households, older adults, and people with disabilities. In this context, the social impact of the disaster centers on how social media communication shapes public risk perception, trust in authorities, and equitable access to relief during emergency response.
In this paper, we present a comprehensive dataset that consolidates multimodal posts (e.g., text, images, videos) from multiple social media platforms to study the societal impact of hurricanes. The dataset is annotated across multiple dimensions (e.g., Humanitarian, Bias, and Information Integrity) to facilitate research on the societal impacts of hurricanes. By providing labeled data, this dataset seeks to support future studies on disaster response such as early detection of false information, identification of vulnerable communities in need of aid, humanitarian response during disasters, and evaluation of public trust in emergency communication.

Existing datasets often focus on hurricanes and other natural disasters that formed several years ago. For instance, many widely used datasets such as CrisisMMD~\cite{crisismmd}, HIM-Twitter~\cite{HIM-Twitter}, and Eyewitness Messages~\cite{eyewitness} focus on the 2017 hurricane season, which is 8 years ago. However, social media was far less developed at the time than it is today, with a much greater number of users and platform diversity than before. As a result, these datasets have limited effectiveness in studying the societal impacts of hurricanes in the current context. In addition, previous datasets predominantly collected posts from a single platform, X (previously known as Twitter), while neglecting other widely used platforms. By excluding posts from diverse platforms, such datasets overlook critical variations in user behavior and communication styles across different social media. This narrow focus fails to capture the broader spectrum of social media activities, leading to datasets that are insufficient for analyzing the comprehensive societal impact of hurricanes. Furthermore, existing datasets either focus exclusively on the textual components of social media posts or treat textual and visual elements independently~\cite{sandy2012,crisismmd,HIM-Twitter,TweetDIS,eyewitness,MEDIC,humaid}. This fragmented treatment overlooks the inherently multimodal nature of social media posts, thereby impeding a comprehensive multimodal understanding of social media content. This gap highlights the need for more comprehensively annotated and up-to-date datasets that incorporate multimodal content from multiple platforms to provide a comprehensive understanding of hurricanes’ societal effects.

Motivated by the identified gaps in existing datasets, we introduce a novel multiplatform and \textbf{M}ultimodal \textbf{A}nnotated Dataset for \textbf{S}ocietal Impact of \textbf{H}urricane (\textbf{MASH}).
Figure~\ref{fig:overview} shows the overview of the MASH dataset and the online platform.
In particular, the MASH dataset consists of 59,607 relevant \textit{social media posts}
collected from three platforms: Reddit, TikTok, and YouTube since they are widely used during hurricanes and provide API for large-scale data collection. All relevant posts are annotated 
along three dimensions: \textit{Humanitarian Classes}, \textit{Bias Classes}, and \textit{Information Integrity Classes} in a multi-modal approach that jointly considers textual and visual content, providing a rich labeled dataset for in-depth analysis. The data collection and annotation framework can also be deployed to other disasters like wildfires and earthquakes.
We also develop an online platform
to facilitate the data exploration and real-time analysis for end users such as researchers, government officials, business owners, and healthcare providers. Specifically, the online platform has the following functions. First, it contains an insight page that can retrieve social media and news media data via customizable keywords to support the investigation of emerging disasters. Second, it contains a highlight page to show some preliminary analysis results (e.g., sentiments, word clouds) of the MASH dataset. Third, it allows users to share observations, interpretations, and community insights about the dataset and the analytical findings.


The development of the MASH dataset is the joint effort of an interdisciplinary research team from information science, computer science, hydrology, and environmental engineering. To the best of our knowledge, MASH is the first large-scale, multimodal, multi-platform, and multi-dimensionally annotated dataset centered on hurricane disasters. Beyond data construction, we further demonstrate the usefulness of MASH through a series of analyses, including correlation analysis, temporal analysis, spatial analysis, baseline model evaluation, and sentiment analysis. We envision that the comprehensiveness of platforms, the multimodality of the data, and the multidimensionality of the annotations will contribute to future studies on disaster response, bias detection, truth discovery, sentiment trajectory analysis, and disaster policy making during natural disasters.



\section{Related Works} \label{RelatedWorks}
With the growth of web-based communication platforms and the ubiquity of the Internet, social media has become a rich resource for exploring the impact of specific events on human society. Recently, several datasets have been collected to study the societal impact of social media posts in the context of hurricanes. Sandy 2012~\cite{sandy2012}, TweetDIS~\cite{TweetDIS}, and NaturalHazards~\cite{NaturalHazards} are datasets that cover tweets during the 2012 Hurricane Season, especially during Hurricane Sandy. In addition, CrisisMMD~\cite{crisismmd}, HIM-Twitter~\cite{HIM-Twitter}, TweetDIS~\cite{TweetDIS}, Eyewitness Messages~\cite{eyewitness}, MEDIC~\cite{MEDIC}, Natural Hazards~\cite{NaturalHazards}, and HumAID~\cite{humaid} primarily capture social media discussions during the 2017 hurricane season, especially during Hurricanes
Harvey, Irma, and Maria. Table~\ref{tab:comparison} shows the comparison between the MASH dataset and prior datasets on hurricanes. The hurricanes that prior datasets focus on are all 8 years ago or even more than 10 years ago. With the progress of society and the development of technology, a single social media platform (X) at that time can no longer well reflect the impact of hurricanes on society. In contrast, in this study, we collected posts from three social media platforms (Reddit, TikTok, and YouTube) and focused on the recent 2024 hurricane season, especially the category 4 Hurricane Helene and category 5 Hurricane Milton. In addition, prior datasets are limited to Humanitarian Class annotation and labeled textual and visual content separately~\cite{sandy2012,crisismmd,HIM-Twitter,TweetDIS,eyewitness,MEDIC,humaid}, which restricts their ability to support comprehensive analyses of hurricanes’ societal impact. In comparison, we annotated social media posts in three dimensions (Humanitarian Class, Bias Class, and Information Integrity Class) by jointly analyzing textual and visual content, enabling more robust and multimodal machine learning models for disaster-related social media analysis. Moreover, previous datasets only contain a limited set of analyses such as temporal analysis, baseline model evaluation, and sentiment analysis~\cite{MEDIC,TweetDIS,HIM-Twitter,sandy2012,NaturalHazards,eyewitness}. On the contrary, we perform systematic analyses of the MASH dataset, including correlation analysis, temporal analysis, spatial analysis, baseline model evaluation, and sentiment analysis.

\section{Data Collection} \label{DataCollection}

The MASH dataset contains social media data that focuses on recent hurricanes, especially the two major hurricanes that hit the United States during the fall of 2024: \textit{Hurricane Helene} and \textit{Hurricane Milton}, to obtain comprehensive information about the societal impact of hurricanes. According to National Hurricane Center,
Hurricane Helene was a Category 4 Storm formed on September 24, 2024 and Hurricane Milton was a Category 5 Storm formed on October 5, 2024. The combined damage from these two consecutive hurricanes is estimated to be more than \$300 billion~\cite{ScienceHurricane}. To ensure timely and relevant coverage of the hurricane-related discussions, the data collection period spans from September 1 to November 30, 2024. This collection window spans approximately one month before the hurricanes and one month after the main impact period, enabling us to capture pre-disaster signals, peak impacts, and recovery-related discussions.

We collected social media posts from three platforms: Reddit, TikTok, and YouTube with a total of 82,095 posts. These three platforms are widely used by diverse demographics and communities as they provide users with different ways to disseminate information in the multi-modality form of text, images, and videos~\cite{TTandYouTube}. 
The social media data collection adopted the Reddit API\footnote{https://www.reddit.com/dev/api}, TikTok Research Tools\footnote{https://developers.tiktok.com/products/research-api}, and YouTube API\footnote{https://developers.google.com/youtube/v3}.
We did not include data from X because the platform discontinued free academic API access in 2023 and replaced it with expensive paid tiers, making large-scale data collection financially infeasible for this project~\cite{murtfeldt2024rip, pehlivan2025can, bisiani2025uktwitnewscor}. We utilized three keywords: \textit{Hurricane}, \textit{Hurricane Helene}, and \textit{Hurricane Milton} to retrieve relevant content.
To ensure the ethics of the research and protect users' privacy, we only collected post information (i.e., post ID, post content, post time, and post location) without users' identity information. Table~\ref{tab:CollectionNum} presents the number of posts collected from three social media platforms. For text-centric platforms like Reddit, we not only collected the title and description of the post but also collected the attached images and videos. For video-centric platforms such as TikTok and YouTube, we collected the video together with the textual data, including title and description. To reduce the cost and workload for further cleaning and annotation tasks, we only collected videos from TikTok and YouTube that are less than five minutes long.


\begin{table}[t]
\centering
\setlength{\tabcolsep}{4pt}
\footnotesize 

\begin{tabular}{ccccc}
\toprule
  & \textbf{Reddit}  & \textbf{TikTok} & \textbf{YouTube} \\
\midrule
Collected Raw Posts    & 12,301  & 67,027           & 2,767           \\
Relevant and Annotated Posts    & 9,928               & 47,915           & 1,764          \\
Avg. \#Words per Annotated Post    &119                & 21           & 18           \\
\#Images on Annotated Posts    & 1,579                 & ---           & ---           \\
\#Videos on Annotated Posts & 878                 & 47,915           & 1,764 \\       
\bottomrule
\end{tabular}
\caption{Distribution of Collected and Annotated Posts.}
\label{tab:CollectionNum}
\end{table}

\section{Data Annotation} \label{DataAnnotation}
The MASH dataset contains annotations for social media posts in three dimensions: \textit{Humanitarian Class}, \textit{Bias Class}, and \textit{Information Integrity Class}. In natural disasters, the main purpose of humanitarian assistance is to save lives, reduce suffering, and rebuild affected communities~\cite{humaid}. For the Humanitarian Class annotations, we defined 7 categories 
to classify the content of posts according to the textual and visual context of the post. The content of social media posts may contain bias that can affect user perception~\cite {MBIB,YifanBiases}. The task of bias class annotation is to identify the presence of defined biases or discriminatory elements in 5 categories within the post. False claims arise in environments when people face a scarcity of needed information~\cite{misinformation, liu2025modality}. The information integrity category assesses the veracity of posts in the face of hurricane disasters, distinguishing between content with verifiable facts and incorrect content. Together, these annotations provide a comprehensive foundation for analyzing the societal impact of hurricanes on
social media posts, enabling research in areas such as detection of false claims, bias, and critical events; physi-social impact forecasting; and vulnerability assessment.
Prior to annotation, we performed a cleaning step to filter out irrelevant posts, ensuring that the subsequent labeling process focuses only on meaningful content. In this section, we introduce the annotation methodology, as well as the label distributions and findings.

\subsection{Annotation Methodology} \label{AnnotationMethodoldy}

To reduce the time and cost of manual annotation, we leverage a Multimodal Large Language Model (MLLM) to support the annotation. We note that a single judgment by the MLLM may contain errors and not be reliable. Therefore, we design a human-MLLM collaborative framework for both data cleaning and annotation, which includes consistency checks based on multiple rounds of MLLM-generated labels and human verification of MLLM-generated labels.

In the first stage, inspired by the work of \citet{wang2023selfconsistency} and \citet{chen2024universal}, the MLLM is queried three times per post to generate three separate labels along with corresponding explanations.
The input to the MLLM consists of task-specific prompts (available in the Appendix Figures~\ref{fig:prompt-clean} to~\ref{fig:prompt-info-inte}) and the full content of each social media post, which includes textual descriptions, and includes Images and videos if they are present. Different from prior datasets (e.g., CrisisMMD~\cite{crisismmd}), which annotate textual and visual content \textit{independently}, we treat each post as a holistic unit and assign labels based on the \textit{joint interpretation} of text and visuals. To protect user privacy, all tagged usernames, starting with “@”, are replaced with “@user”. This multi-modal input enables the MLLM to generate labels based on a comprehensive understanding of the post across all available modalities. To measure the consistency of the MLLM’s outputs, we compute the entropy of the three MLLM-generated-labels. A zero entropy score, indicating all three predictions are identical, reflects a high degree of confidence in the MLLM’s judgment. In such cases, the label is considered high-confidence and directly adopted. 

On the other hand, if the label entropy computed in the first stage is non-zero, indicating that the MLLM exhibits uncertainty regarding the task, we proceed to the second stage. In the second stage, we provide the MLLM with all three sets of labels and corresponding reasons generated in the first stage. The MLLM is prompted (available in the Appendix Figure~\ref{fig:prompt-second}) to compare, analyze, and determine which label and justification are more reasonable. This setup encourages the model to perform consistency resolution and preference selection, aligning with recent work demonstrating that MLLM benefits from self-consistency~\cite{wang2023selfconsistency} and iterative self-refinement strategies~\cite{madaan2023selfrefine, shinn2023reflexion} to improve label reliability and reason quality. The MLLM subsequently generates three new sets of labels and reasons, from which the new entropy is computed. If the new entropy value is zero, the new label is also considered as a high-confidence label and accepted accordingly. If the new entropy remains non-zero, indicating persistent uncertainty, the post is escalated to the third stage for human annotation. In this stage, human annotators from diverse disciplines in our team collaboratively review the post and determine the appropriate label. Through open discussion, annotators exchange perspectives and reach consensus on the final label. This human annotation process ensures that ambiguous cases are carefully reviewed. 

Furthermore, we conduct two separate human verification tasks to evaluate the reliability of the MLLM-generated labels.
Firstly, for the data cleaning task, we randomly sample 500 posts from each social media platform (1,500 posts in total) from the raw collected dataset, and three annotators in our team independently verify whether each post is relevant to the disaster.
Secondly, for the annotation task, we randomly sample another 500 posts from each platform (1,500 posts in total) from the cleaned relevant dataset. For every sampled post, three annotators independently verify all labels generated by the MLLM, including the labels in Humanitarian, Bias, and Information Integrity Classes.
The verification results for both tasks show high consistency and are reported in Section Annotation Consistency. We used Gemini-2.0-Flash as the MLLM in both data cleaning and annotation tasks because of its cost efficiency and advanced capabilities over multimodal inputs, which is suitable for our large-scale dataset~\cite{jegham2025visual, hirosawa2024comparative, team2023gemini, team2024gemini}.

\subsection{Data Cleaning}\label{DataClean}

We observe that the collected raw social media posts contain irrelevant and off-topic posts, which could reduce the quality of the dataset. For example, using the keyword “hurricane” in API search may retrieve posts about the Miami Hurricanes football team or Carolina Hurricanes hockey team, which are not related to the hurricane disaster. 
Additionally, some posts use hurricane-related hashtags solely for promotional purposes, such as adding \#hurricane to increase exposure and attract more attention to their advertisement. These practices often result in posts with irrelevant content being included in the dataset.
To address this issue, we implement the human-MLLM collaborative annotation framework
to keep posts related to actual North American hurricane disasters (particularly Hurricane Helene and Hurricane Milton) and exclude content that literally mentions ``hurricane'' but is not relevant to an actual disaster, such as sports, entertainment, and metaphors. The detailed prompt can be found in the Appendix Figure~\ref{fig:prompt-clean}. Table~\ref{tab:CollectionNum} presents the distribution of relevant posts for each social media platform after cleaning. A total of 59,607 posts were identified as relevant and subsequently annotated across the three dimensions. The results in Table~\ref{tab:annotation_consistency} show high agreement both among human annotators and between human and the MLLM across all platforms, indicating the reliability of data cleaning.

\subsection{Humanitarian Class Annotation}
The purpose of the Humanitarian Class annotation is to classify social media posts based on their themes before, during, and after the hurricane disaster, facilitating a semantic understanding of the data sample. 
This classification also supports the analysis of public communication and information dissemination during natural disasters like hurricanes. Based on previous datasets~\cite{crisismmd,eyewitness,gebru2021datasheets,HIM-Twitter,sandy2012,TweetDIS,humaid}, we classify the social media data samples into 7 categories: \textbf{Casualty} (Cslt), \textbf{Evacuation} (Evac), \textbf{Damage} (Dmg), \textbf{Advice} (Advc), \textbf{Request} (Rqst), \textbf{Assistance} (Aid), \textbf{Recovery} (Rcv). The detailed definition of these classes is available in the Appendix Table~\ref{tab:def-human}. If the content of the post does not fit into any of these categories we defined, it is classified as \textbf{Other Useful Information} (OUI).

We utilized the MLLM-human collaborative framework introduced in Section~\ref{AnnotationMethodoldy} to annotate the Humanitarian Class of each post. The prompt is presented in the Appendix Figure~\ref{fig:prompt-human-bias} and the human verification is presented in Section Annotation Consistency. Table~\ref{tab:DistributionHumanitarianClasses} reports the distribution of Humanitarian Classes of posts from each social media platform, with the largest value in each platform highlights in bold and the second-largest value indicates with underline. We note that a social media post may contain information that belongs to multiple Humanitarian Classes. For example, a post describing disaster situations might mention both casualties and damage of infrastructure. Consequently, the cumulative percentages of posts across all Humanitarian Classes exceed 100\% in Table~\ref{tab:DistributionHumanitarianClasses}. From Table~\ref{tab:DistributionHumanitarianClasses} we observe that the categories Damage and Recovery have high prevalence across all platforms, 
suggesting that users frequently upload and discuss content related to the physical destruction caused by hurricanes and the subsequent recovery efforts.


\begin{table}[t]
\footnotesize
    \centering
    \setlength{\tabcolsep}{3pt}

    \begin{tabular}{ccccc}
    \toprule
      & \textbf{Reddit} & \textbf{TikTok} & \textbf{YouTube} \\
    \midrule
\textbf{Casualty}   & 891 (9\%) & 4,788 (10\%)           & 310 (18\%)           \\ 
\textbf{Evacuation}    & 2,475 (25\%)   & 15,009 (31\%)           & 728 (41\%)           \\ 
\textbf{Damage}    & \textbf{4,340 (44\%)}  & \underline{22,898 (48\%)}           & \textbf{1,364 (77\%)}           \\ 
\textbf{Advice}    & 2,079 (21\%)   & 14,461 (30\%)           & 334 (19\%)           \\ 
\textbf{Request}    & 2,750 (27\%)   & 8,804 (18\%)           & 416 (24\%)           \\ 
\textbf{Assistance}    & 3,653 (37\%)   & \textbf{23,940 (50\%)}           & 1,015 (58\%)           \\ 
\textbf{Recovery}    & \underline{4,280 (43\%)}  & 22,437 (47\%)           & \underline{1,122 (64\%)}           \\  
\midrule
\textbf{Other Useful Info}   & 2,511 (25\%)   & 8,991 (19\%)           & 145 (8\%)           \\ 
    \bottomrule
    \end{tabular}
\captionof{table}{Distribution of Humanitarian Classes}

    \label{tab:DistributionHumanitarianClasses}
\end{table}

\subsection{Bias Class Annotation}
 Bias Class annotation aims to identify the existence of bias in social media posts. \citet{gu2021social} and \citet{lifang2020effect} point out that social media users tend to express more polarized and biased opinions during emergencies like natural disasters because of tension and anxiety. Based on previous studies~\cite{MBIB, YifanBiases, SystematicBias}, we define 5 post-level bias types, \textbf{Linguistic Bias} (LB), \textbf{Political Bias} (PB), \textbf{Gender Bias} (GB), \textbf{Hate Speech} (HS), \textbf{Racial Bias} (RB). 
 The purpose of the Bias Class annotation is to characterize how biased or discriminatory narratives emerge during disasters, enabling researchers to study their patterns, impacts, and potential consequences for different communities.
 The detailed definition of these classes can be found in the Appendix Table~\ref{tab:def-bias}. If the content of the post does not fit into any of the biases we defined, it is classified as \textbf{Undefined Bias} (UB).


Since the only difference between the Bias and Humanitarian annotations is the definitions of categories, we also adopt the human-MLLM collaborative framework introduced in Section~\ref{AnnotationMethodoldy} to annotate Bias Classes. The prompt is presented in the Appendix Figure~\ref{fig:prompt-human-bias} and the human verification is presented in Section Annotation Consistency.
Table~\ref{tab:DistributionBiasClasses} reports the distribution of Bias Classes of posts from each social media platform with the largest value in each platform highlights in bold and the second-largest value indicates with underline. 
As illustrated in Table~\ref{tab:DistributionBiasClasses}, the majority of posts are free from defined bias, while a subset of posts exhibits one or multiple types of bias.
Table~\ref{tab:DistributionBiasClasses} demonstrates that all three platforms exhibit a relatively high ratio of posts classified as Linguistic Bias and Political Bias. This trend may be attributed to the strong emotional reactions triggered by hurricane disasters, which often prompt people to express their views in more intense or confrontational language, thus leading to linguistic bias. Additionally, posts about government relief efforts often reflect political leanings, especially given the close timing of Hurricanes Helene and Milton and the 2024 U.S. presidential election. This overlap in time could exacerbate political discussions and criticisms, further exacerbating the presence of political bias in these posts.
\begin{table}[t]
\footnotesize
    \centering
    \setlength{\tabcolsep}{4pt}

    \begin{tabular}{ccccc}
    \toprule
  & \textbf{Reddit} & \textbf{TikTok} & \textbf{YouTube} \\
\midrule
\textbf{Linguistic Bias}    & \textbf{1,914 (19\%)}  & \textbf{13,371 (28\%)}           & \textbf{241 (14\%)}           \\ 
\textbf{Political Bias}    & \underline{1,798 (18\%)}   & \underline{5,405 (11\%)}           & \underline{238 (13\%)}        \\ 
\textbf{Gender Bias}    & 190 (2\%)    & 2,891 (6\%)           & 55 (3\%)           \\ 
\textbf{Hate Speech}    & 106 (1\%)    & 1,249 (3\%)           & 17 (1\%)           \\ 
\textbf{Racial Bias}    & 144 (1\%)   & 1,444 (3\%)           & 27 (2\%)           \\ 
\midrule
\textbf{Undefined Bias}     & 7,141 (72\%)   & 31,508 (66\%)           & 1,334 (76\%)           \\ 
    \bottomrule
    \end{tabular}
    \captionof{table}{Distribution of Bias Classes}

\label{tab:DistributionBiasClasses}
\end{table}

\subsection{Information Integrity Class Annotation}

The primary goal of the Information Integrity Class annotation task is to determine whether the content described in a post is factual or constitutes false arguments.
The Information Integrity Class includes three distinct labels: \textit{True Information}, \textit{False Information}, and \textit{Unverifiable Information}. The inclusion of the Unverifiable Information label accounts for posts that share personal or subjective opinions, which cannot be objectively verified. Different from Humanitarian and Bias class annotation, information integrity annotation raises unique challenges as it requires factual verification that cannot be reliably performed by the MLLM alone. Additionally, the MLLM we adopted in the annotation framework, Gemini-2.0-flash, does not contain up-to-date knowledge beyond August 2024, making it difficult to verify the factuality of posts published between September to November 2024. To address this limitation, we enabled Gemini’s online function (i.e., Grounding with Google Search\footnote{https://ai.google.dev/gemini-api/docs/grounding})
to support truth discovery.
This functionality allows the model to retrieve relevant external content based on the content of the input post, such as government announcements, truth discovery
websites, and news media reports. The retrieved information serves as an external source of evidence, enabling the MLLM to assess the factuality of the post based on real-time and authoritative content. A post is labeled as \textit{True Information} only when its core claims are explicitly supported by consistent external evidence. Posts are labeled as \textit{False Information} when authoritative sources directly contradict the claims. When retrieved evidence is absent, inconsistent, or insufficient to support a clear judgment, the post is labeled as \textit{Unverifiable Information}. The prompt for the Information Integrity Class annotation is present in the Appendix Figure~\ref{fig:prompt-info-inte} and the human verification is presented in Section Annotation Consistency. Due to the strict rate limits imposed by the Grounding with Google Search, we limited the MLLM to a single round. 

Table~\ref{tab:FakeNewsDistribution} reports the distribution of Information Integrity Class for each social media. Across the three platforms, the percentage of posts labeled as factual information is consistently around 75\%, demonstrating a notable level of uniformity. The proportion of posts labeled as false claims
on TikTok and YouTube is consistent around 18\%. In contrast, Reddit exhibits a lower proportion of false claims
and accompanied by a corresponding increase in the proportion of posts categorized as unverifiable information. This pattern suggests that Reddit may contain fewer false claims and more content that lacks sufficient evidence for verification.

\begin{table}[t]
\footnotesize
\centering

\setlength{\tabcolsep}{4pt}
\begin{tabular}{ccccc}

\toprule
  & \textbf{Reddit}  & \textbf{TikTok} & \textbf{YouTube} \\
\midrule
\textbf{True Info}     & 7,304 (74\%)   & 33,440 (70\%)  & 1,352 (77\%)                      \\ 
\textbf{False Info}    & 746 (7\%)  & 8,682 (18\%)  & 324 (18\%)                      \\  
\textbf{Unverifiable Info}     & 1,878 (19\%)    & 5,793 (12\%)  & 88 (5\%)                     \\  
\bottomrule
\end{tabular}
\caption{Distribution of Information Integrity Classes}

\label{tab:FakeNewsDistribution}
\end{table}

\subsection{Annotation Consistency} \label{anno_consistency}

\begin{table*}[t]
\centering
\setlength{\tabcolsep}{1pt}
\footnotesize
\begin{tabularx}{\textwidth}{
    c|
    >{\centering\arraybackslash}X |
    >{\centering\arraybackslash}X |
    >{\centering\arraybackslash}X ||
    >{\centering\arraybackslash}X |
    >{\centering\arraybackslash}X |
    >{\centering\arraybackslash}X
}
\toprule[1.5pt]
\multirow{3}{*}{\textbf{Category}} 
& \multicolumn{3}{c||}{\textbf{Inter-Annotator Consistency}} 
& \multicolumn{3}{c}{\textbf{Human-MLLM Consistency}} \\
\cmidrule(lr{0.75em}){2-4} \cmidrule(lr{0.75em}){5-7}
& Reddit & TikTok & YouTube & Reddit & TikTok & YouTube \\
& (Acc / Fleiss' $\kappa$)  & (Acc / Fleiss' $\kappa$) & (Acc / Fleiss' $\kappa$) & (Acc / Cohen's $\kappa$) & (Acc / Cohen's $\kappa$) & (Acc / Cohen's $\kappa$)  \\
\midrule
\textbf{Clean}              & $0.93 \mid 0.87$ & $0.99 \mid 0.98$ & $0.97 \mid 0.78$ & $0.99 \mid 0.99$ &  $1.00 \mid 1.00$ & $0.99 \mid 0.84$ \\
\midrule
\textbf{Casualty}              & $0.93 \mid 0.81$ & $0.98 \mid 0.92$ & $0.97 \mid 0.92$ & $0.98 \mid 0.92$ &  $0.99 \mid 0.94$ & $0.98 \mid 0.94$ \\
\textbf{Evacuation}              & $0.93 \mid 0.88$ & $0.94 \mid 0.88$ & $0.97 \mid 0.95$ & $0.97 \mid 0.91$ &  $0.96 \mid 0.90$ & $0.98 \mid 0.95$ \\
\textbf{Damage}               & $0.95 \mid 0.94$ & $0.96 \mid 0.94$ & $0.96 \mid 0.93$ & $0.98 \mid 0.96$ & $0.98 \mid 0.95$ & $0.97 \mid 0.92$ \\
\textbf{Advice}              & $0.92 \mid 0.84$ & $0.96 \mid 0.94$ & $0.97 \mid 0.93$ & $0.98 \mid 0.94$ & $0.99 \mid 0.96$ & $0.98 \mid 0.95$ \\
\textbf{Request}              & $0.93 \mid 0.87$ & $0.97 \mid 0.92$ & $0.97 \mid 0.93$ & $0.98 \mid 0.93$ &  $0.99 \mid 0.97$ & $0.98 \mid 0.95$ \\
\textbf{Assistance}               & $0.94 \mid 0.90$ & $0.95 \mid 0.93$ & $0.97 \mid 0.96$ & $0.97 \mid 0.94$ & $0.97 \mid 0.94$ & $0.98 \mid 0.97$ \\
\textbf{Recovery}               & $0.93 \mid 0.91$ & $0.96 \mid 0.94$ & $0.96 \mid 0.95$ & $0.97 \mid 0.93$ & $0.97 \mid 0.94$ & $0.98 \mid 0.96$ \\
\midrule
\textbf{Linguistic Bias}                & $0.93 \mid 0.85$ & $0.94 \mid 0.89$ & $0.97 \mid 0.91$ & $0.99 \mid 0.97$ &  $0.97 \mid 0.91$ & $0.99 \mid 0.94$ \\
\textbf{Political Bias}                & $0.93 \mid 0.85$ & $0.99 \mid 0.95$ & $0.98 \mid 0.94$ & $0.99 \mid 0.96$ &  $0.99 \mid 0.99$& $0.99 \mid 0.93$ \\
\textbf{Gender Bias}                & $0.99 \mid 0.82$ & $0.99 \mid 0.90$ & $0.99 \mid 0.87$ & $0.99 \mid 0.91$ & 
$0.99 \mid  0.92$ & $0.99 \mid 0.79$ \\
\textbf{Hate Speech}                & $0.99 \mid 0.66$ & $0.99 \mid 0.95$ & $0.99 \mid 0.83$ & $0.99 \mid 0.86$ &  $1.00 \mid 1.00$ & $0.99 \mid 0.86$ \\
\textbf{Racial Bias}                & $0.99 \mid 0.79$ & $0.99 \mid 0.86$ & $0.99 \mid 0.92$ & $1.00 \mid 1.00$ &  $0.99 \mid 0.94$ & $0.99 \mid 0.95$ \\
\midrule
\textbf{Info Integrity}           & $0.91 \mid 0.85$ & $0.88 \mid 0.81$ & $0.95 \mid 0.91$ & $0.95 \mid 0.88$ &  $0.92 \mid 0.82$ & $0.97 \mid 0.93$ \\
\bottomrule[1.5pt]
\end{tabularx}
\caption{Inter-Annotator and Human–MLLM Consistency by each Category}

\label{tab:annotation_consistency}
\end{table*}

Similar to \citet{manakul2023selfcheckgpt}, we invite three human annotators from different disciplines to manually verify the reliability of the MLLM-generated outputs.
We design two tasks to verify the reliability: data cleaning verification and data annotation verification. In the data cleaning verification, for each platform, three annotators independently review 500 randomly sampled posts from the raw collected dataset (1,500 posts in total) to confirm whether the posts are related to the hurricane disaster. In the annotation verification, for each platform, we randomly sample another 500 posts (also 1,500 posts in total) from the cleaned relevant dataset. For each sampled post, annotators verify all MLLM-assigned labels across the Humanitarian, Bias, and Information Integrity Classes.
For the Information Integrity Class, we focus on true and false claims, thereby reducing all evaluation tasks to binary classification. To evaluate the reliability, we measure both inter-annotator agreement among the human annotators and the agreement between the human majority vote and the output generated by MLLM. We adopt Accuracy and Fleiss’ Kappa Score for inter-annotator evaluation. Accuracy calculates the proportion of samples that are completely consistent among all annotators, while Fleiss’ Kappa Score calculates the overall consistency after random consistency correction between annotators. Fleiss’ Kappa is suitable for measuring agreement among three or more raters, making it appropriate for evaluating inter-annotator reliability~\cite{fleiss1971measuring}.
Subsequently, we apply majority voting to aggregate human annotations.
We compute Accuracy and Cohen’s Kappa to evaluate the consistency between human consensus and MLLM's annotations. Different from Fleiss’ Kappa, Cohen’s Kappa quantifies agreement between two raters by correcting for the proportion of agreement that would be expected purely by chance, making it more appropriate for pairwise agreement assessment compared to Fleiss’ Kappa~\cite{cohen1960coefficient}.
Table~\ref{tab:annotation_consistency} shows the consistency of the annotations among the three annotators and the consistency between the MLLMs' annotations and the human consensus. We find that human inter-annotator agreement is strong, with Fleiss’ Kappa consistently near or above 0.8. Similarly, the agreement between MLLM's label and human consensus is also high, with Cohen’s Kappa above 0.8 and accuracy exceeding 0.9, which reflects the reliability of MLLM's annotations. We observe that categories such as Hate Speech and Gender Bias show relatively lower agreement. A plausible explanation is their highly imbalanced label distributions, which make the metrics more sensitive to small annotation disagreements.

\subsection{Dataset Availability}
We follow the  FAIR  Data  Principles~\cite{fair} to release the MASH dataset by only providing the IDs and annotations of the posts. This approach not only ensures that the data obtained from IDs is up-to-date, but also protects user privacy by preventing the disclosure of users' personal identifiable information. The MASH dataset is findable with a unique Digital Object Identifier (DOI): https://doi.org/10.5281/zenodo.17682231. 
The MASH  dataset is reusable under the
Creative Commons Attribution 4.0 (CC BY 4.0) license, with a README file in the repository to explain the proper usage.

\section{Preliminary Analyses}\label{DataAnalysis}

\label{PreliminaryAnalyses}

In this section, we comprehensively analyze the collected social media data and their annotations across various dimensions. Specifically, we perform Correlation Analysis, Temporal Analysis, Spatial Analysis, Sentiment Analysis, Multi-modal Baseline Model Evaluation, and Text-based Baseline Model Evaluation.

\subsection{Correlation Analysis}

\begin{figure}[t]
    \centering
    \begin{minipage}{0.48\textwidth} 
        \centering
        \includegraphics[width=\textwidth]{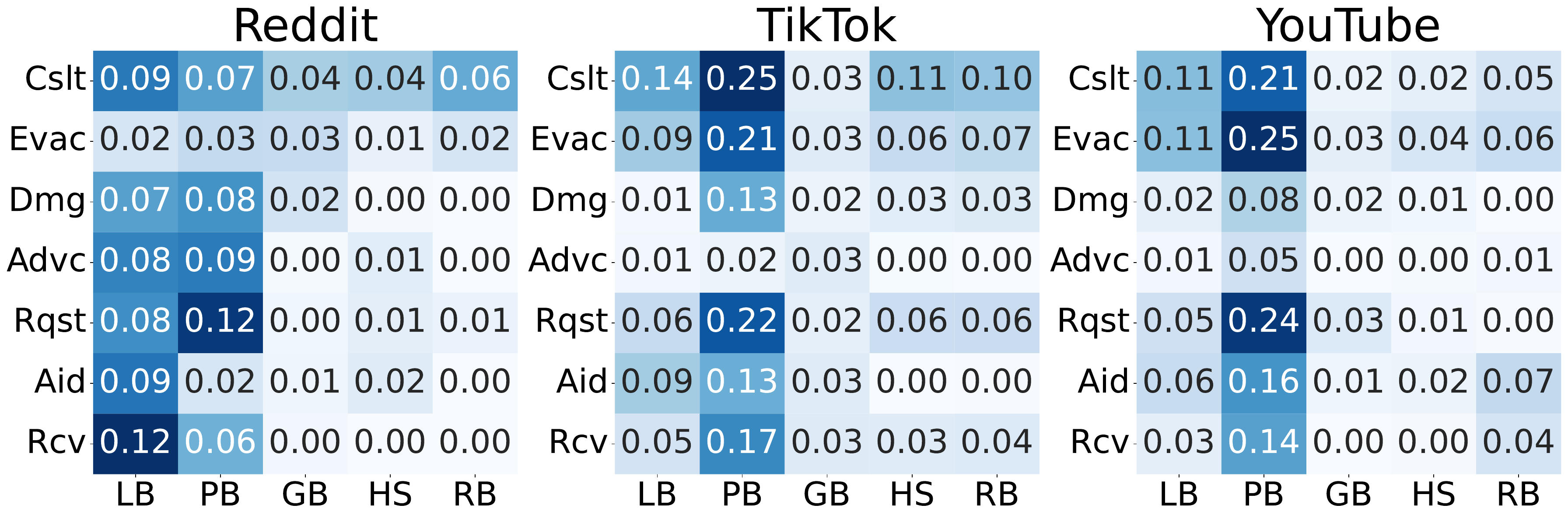} 
        \caption{Humanitarian Class vs. Bias Class Correlation} 
        \label{fig:corr_human_bias} 
    \end{minipage}
\end{figure}

\begin{figure}[t]
\centering
\footnotesize

\begin{minipage}{0.35\columnwidth}
    \centering
    \includegraphics[width=\linewidth]{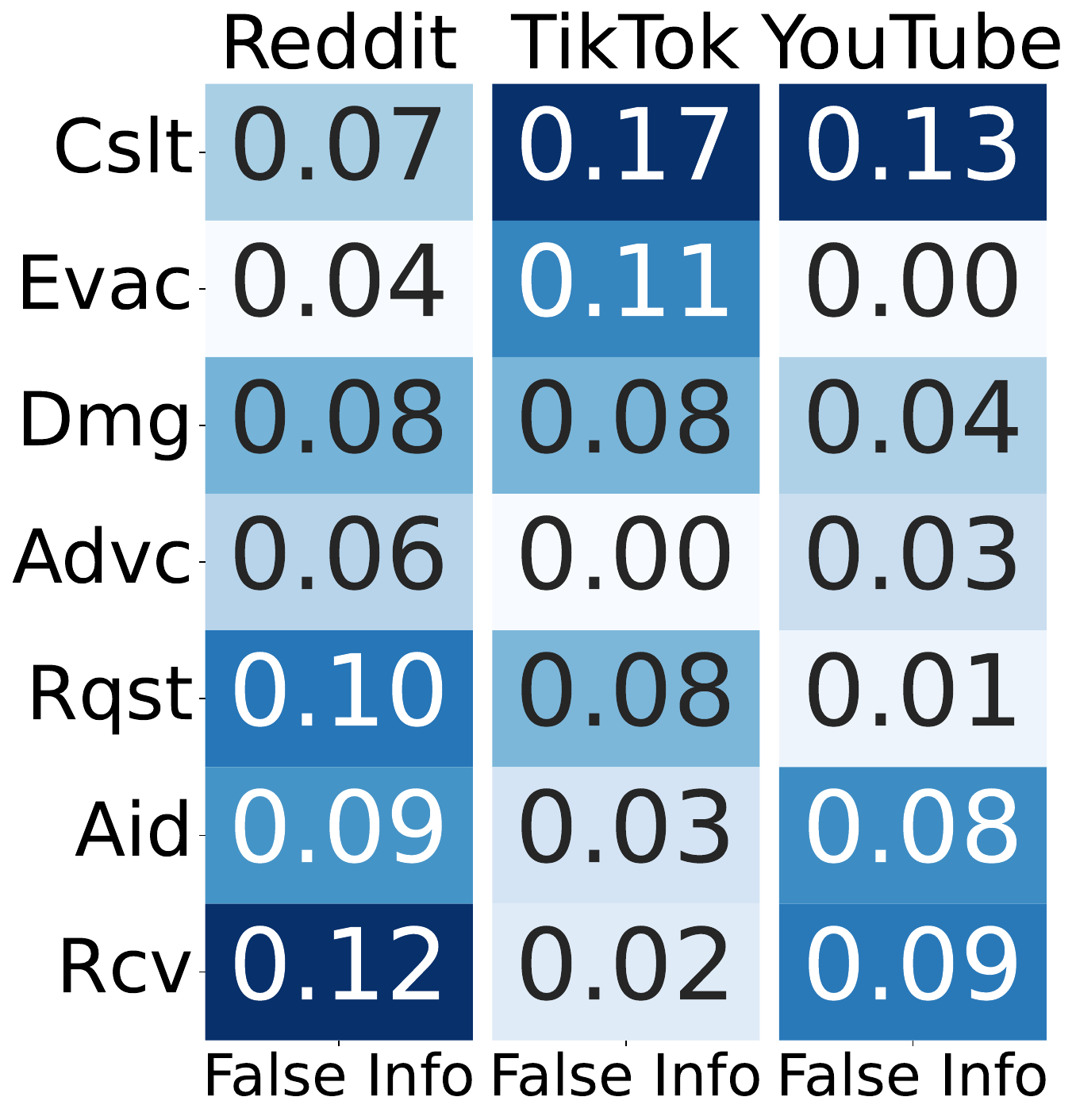}
    \textbf{(a) False Info vs. Humanitarian}
    \label{fig:fake_news_humanitarian_corr}
\end{minipage}
\hspace{0.02\textwidth} 
\begin{minipage}{0.35\columnwidth}
    \centering
    \includegraphics[width=\linewidth]{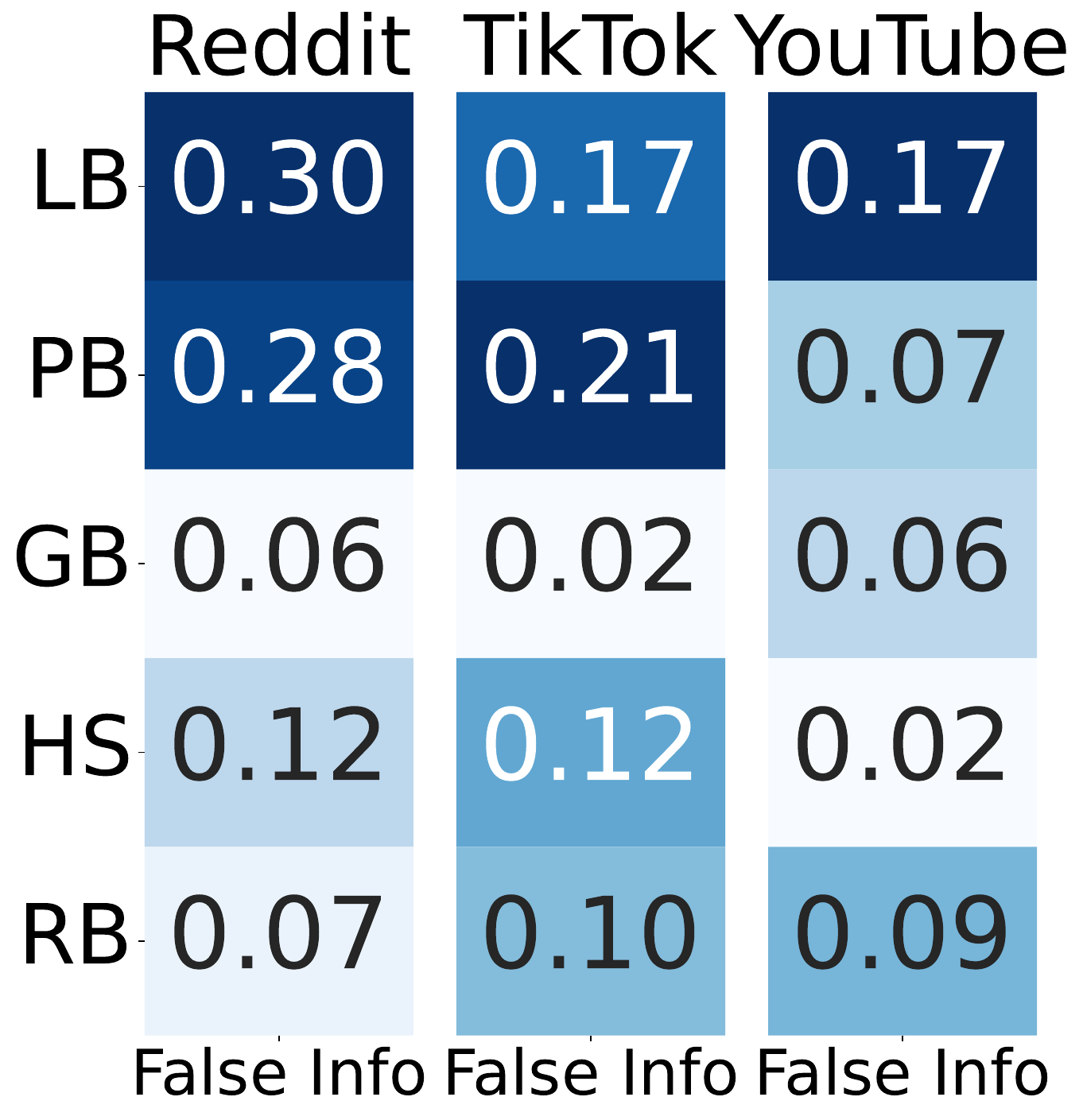}
    \textbf{(b) False Info vs. Bias}
    \label{fig:fake_news_bias_corr}
\end{minipage}

\caption{False Info vs. Humanitarian and Bias Classes.}
\label{fig:corr_fake_humanitarian_bias}
\end{figure}

In correlation analysis, we calculate and analyze the correlations across different annotation dimensions.
In particular, we focus on understanding the connections between annotated labels in different dimensions to uncover potential patterns and interactions within the dataset. We utilize Cramer’s $V$ to measure the correlations between the various annotated labels, where higher values indicate stronger correlations. Specifically, we calculate the correlation based on the co-occurrence of multiple labels in the same post, rather than relying on the overall distribution of posts. This approach ensures that correlation reflects the actual relationship between labels and avoids the bias caused by comparing a large number of posts in different categories~\cite{YifanBiases}. Figure~\ref{fig:corr_human_bias} indicates the correlation relationship between Humanitarian Classes and Bias Classes. We observe that the Request category is strongly associated with Political Bias across all platforms. One possible reason is that users may criticize government responses or call for policy-level actions to support victims. Similarly, the Casualty tends to correlate with both Political and Linguistic Bias, implying that posts discussing human loss often incorporate emotional and polarized expression. In contrast, bias categories such as Gender Bias, Hate Speech, and Racial Bias demonstrate little correlation with humanitarian classes, likely due to their relatively low prevalence in the dataset.

Figure~\ref{fig:corr_fake_humanitarian_bias} (a) illustrates the correlations between false information and the various Humanitarian Classes. We observe that posts related to the Casualty class exhibit the strongest correlation with false information across all platforms. This highlights the prevalence of false information surrounding death and injury reports in the aftermath of disasters.
Figure~\ref{fig:corr_fake_humanitarian_bias} (b) presents the correlations between false information and the Bias Classes. We observe that Linguistic and Political Biases exhibit the strongest correlations with false information across platforms, particularly on Reddit. This points out that false information is often communicated through subjective or ideologically driven language.
The above observations also suggest that the presence of Linguistic and Political Bias can serve as an indicator for identifying false information in social media discussions. In contrast, Gender Bias shows consistently low correlation with false content across all platforms, indicating that gender-related content is largely absent from false information in the disaster context.

\begin{figure*}[t]
    \centering
        \includegraphics[width=\textwidth]{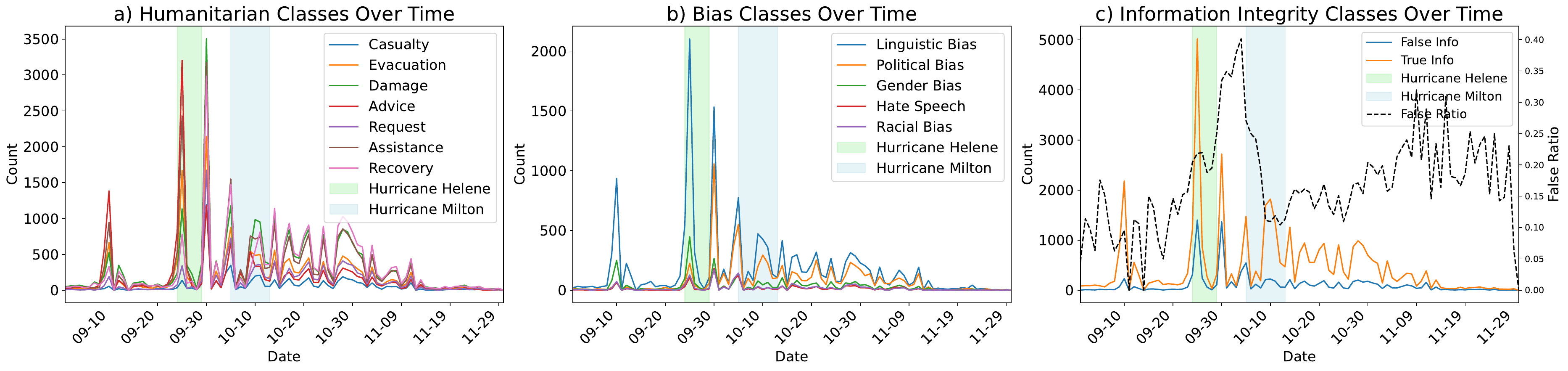}
        \caption{Temporal Trend of Humanitarian, Bias, and Information Integrity Classes}
        \label{fig:temporal}
\end{figure*}

\subsection{Temporal Analysis}\label{temporal}




We conduct a temporal analysis of the social media posts, focusing on the distribution of annotation labels over time. Figure \ref{fig:temporal} illustrates the time series distributions of Humanitarian Classes, Bias Classes, and Information Integrity Classes respectively. We observe a significant increase in the number of posts during the periods when Hurricane Helene and Hurricane Milton occurred, reflecting an increase in user activity in response to these events.

For the Humanitarian Classes, the categories Request and Advice are most prominent during the early stages of the hurricane, reflecting how users turn to social media to seek and offer help in anticipation of the disaster. As the hurricane strikes and its impact becomes visible, posts increasingly shift toward the Damage category, capturing reports of destruction and loss. In the aftermath, the Recovery category becomes more dominant, indicating a collective focus on rebuilding efforts.

For the Bias categories analysis, linguistic bias and political bias account for the majority of biased content. Notably, the distribution of both biases increases significantly during hurricanes. The increase in linguistic bias reflects that public discourse becomes more intense during disaster events, as individuals express stronger emotions in response to the crisis. Similarly, the increase in political bias is often driven by criticism of government relief efforts as public scrutiny of relief measures grows. These findings suggest that the impacts of hurricanes are not limited to physical dimensions, such as individual casualties and damage to infrastructure. Hurricanes also affect public discourse, highlighting the broad societal impact.

For Information Integrity Classes, the amount of both true information and false information increases significantly during the hurricanes. We observe that during the period of Hurricane Helene and Hurricane Milton, the proportion of false claims was notably high, peaking at around 40\%. In the aftermath, this proportion dropped and stabilized at around 20\%. This increase shows that as the level of information dissemination increases, the amount of true and false claims also increases. This increase also highlights the urgent need to verify facts during disasters, because the urgency of sharing information during disasters often leads to the spread of false claims. The spread of false claims could further cause public tension about natural disasters.

\subsection{Spatial Analysis}\label{spatial}
\begin{figure}[t]
    \centering
    \begin{minipage}{0.49\textwidth}
        \centering
        \includegraphics[width=\textwidth]{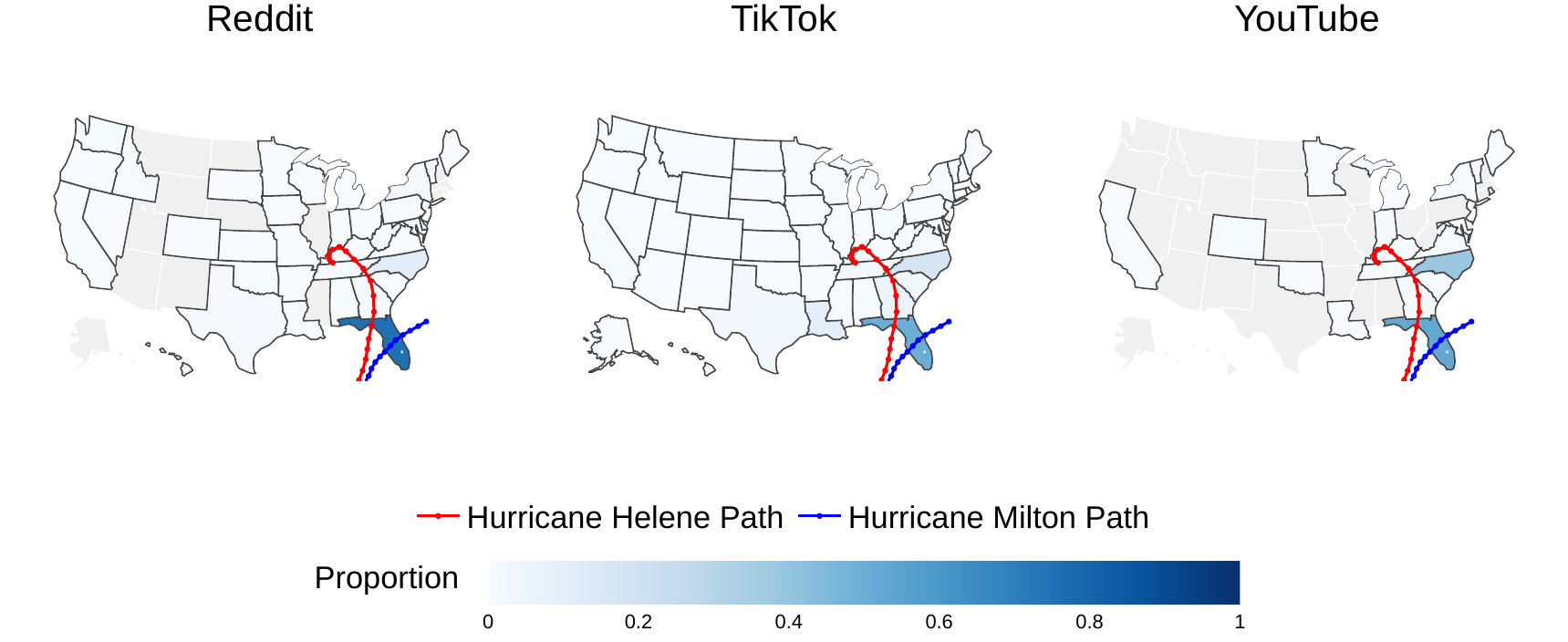}
        \caption{Spatial Distribution of Posts Across U.S. States}
        \label{fig:geo}
    \end{minipage}
\end{figure}

To examine the geographic distribution of posts in MASH dataset, we conduct a spatial analysis based on posts' location information. We employ two methods to extract geographic locations: (1) if a post explicitly discloses its location upon publication, we directly use this metadata; (2) if explicit location data is unavailable, we extract state-level information based on textual mentions of U.S. states within the post content. Posts for which neither method yields valid location information are excluded from this analysis.
Figure~\ref{fig:geo} shows the state-level distribution of geolocated posts, reported as the proportion of posts located in each U.S. state.
Darker shades indicate a higher proportion of posts originating from each state. We observe that a substantial majority of geolocated posts are concentrated in Florida and North Carolina, which aligns with the primary trajectory of Hurricane Helene and Hurricane Miltion captured in the dataset. This pattern highlights the spatial relevance of the social media posts collected in our dataset.

\subsection{Sentiment Analysis}\label{sentiment}

\begin{figure}[t]
    \centering
    \begin{minipage}{0.45\textwidth} 
        \centering
        \includegraphics[width=\textwidth]{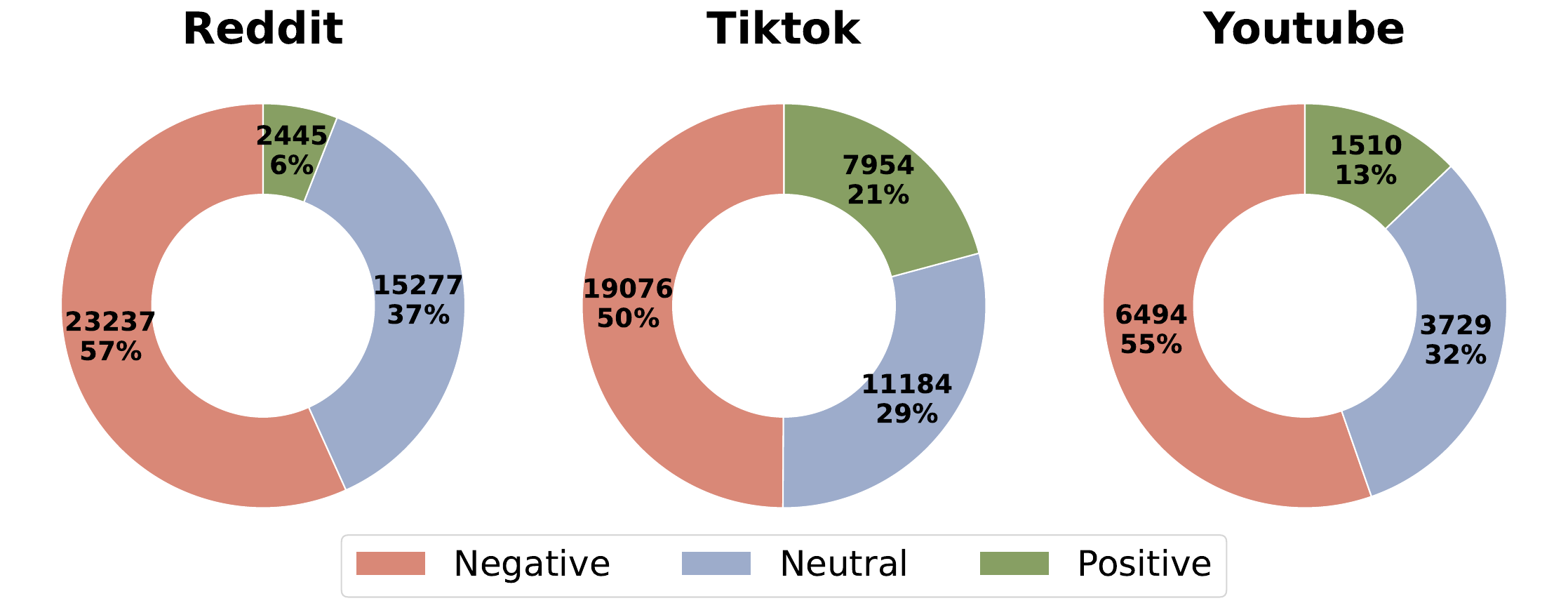} 
        \caption{Sentiment Spectrum of Social Media Platform} 
        \label{fig:sentiment} 
    \end{minipage}
\end{figure}

\begin{table*}[t]
\centering

\footnotesize

\begin{tabularx}{\textwidth}{c|>{\centering\arraybackslash}X|>{\centering\arraybackslash}X||
                                    >{\centering\arraybackslash}X|>{\centering\arraybackslash}X||
                                    >{\centering\arraybackslash}X|>{\centering\arraybackslash}X}
\toprule[1.5pt]
\multirow{2}{*}{\textbf{Category}} 
    & \multicolumn{2}{c||}{\textbf{Phi-4-Multimodal-Instruct}} 
    & \multicolumn{2}{c||}{\textbf{LLaVA-Onevision-7B}}
    & \multicolumn{2}{c}{\textbf{Qwen2.5-VL-7B-Instruct}} 
    \\ 
\cmidrule(lr){2-3} 
\cmidrule(lr){4-5} 
\cmidrule(lr){6-7}
 & Accuracy & Macro F1 
 & Accuracy & Macro F1 
 & Accuracy & Macro F1 \\
\midrule
Casualty             & 0.9070 & 0.5658 &  0.8730 &  0.6117  & 0.9110  & 0.5890\\ 
Evacuation             & 0.7113 & 0.4997 &   0.6475 &  0.5780 & 0.7116  & 0.4719 \\ 
Damage             & 0.5848 & 0.4794 &   0.6809 & 0.6668 & 0.6090  & 0.5218\\ 
Advice           & 0.7364 & 0.5141 & 0.7001  & 0.6052 & 0.7410  & 0.5104\\ 
Request              & 0.8130 & 0.5917 &  0.6915  & 0.5829 & 0.8176  & 0.6022 \\ 
Assistance            & 0.6075 & 0.5214 &  0.6675  & 0.6625  & 0.6089  & 0.5221\\ 
Recovery              & 0.5934 & 0.4815 &  0.6904  & 0.6808 & 0.5980  & 0.4868\\ 
\midrule
Linguistic Bias & 0.7579 & 0.5518 &    0.7239  & 0.5804 & 0.7627  & 0.5428\\  
Political Bias  & 0.9026 & 0.7108 &  0.8695  & 0.7066 & 0.9081  &  0.6919\\  
Gender Bias     & 0.9368 & 0.5281 &  0.9102  & 0.5457 & 0.9461  &  0.5230\\ 
Hate Speech     & 0.9721 & 0.5802 &   0.9475  & 0.5725 & 0.9754  & 0.5580\\  
Racial Bias     & 0.9657 & 0.5916  &  0.9241  & 0.5646  & 0.9726  & 0.5631\\  
\midrule
True/False Information       & 0.5892 & 0.5257 &  0.8353  & 0.5557 & 0.4472  & 0.4449\\ 
\bottomrule[1.5pt]
\end{tabularx}
\caption{Accuracy and Macro F1 for Different Multi-modal Baseline Models in Zero-shot Setting}

\label{tab:baseline}
\end{table*}

We implement sentiment analysis on relevant posts from three social media platforms to examine the kinds of emotions expressed in the content. Following \citet{zhang-etal-2024-sentiment}, we adopt their prompting strategy to guide the MLLM in analyzing the sentiment of each post in our dataset. \citet{zhang-etal-2024-sentiment} demonstrate that the MLLM achieves competitive performance in sentiment classification tasks, especially under few-shot or zero-shot settings, through a comprehensive test across 26 datasets. Therefore, we adopt the method because of the similarity of our task to the tested sentiment analysis task. Specifically, we let the MLLM consider the multimodal content of the post and classify the sentiment of the post into three categories: \textit{Positive}, \textit{Negative}, and \textit{Neutral}. Figure~\ref{fig:sentiment} illustrates the distribution of sentiments across all platforms. We observe that the majority of posts showed negative or neutral sentiment, with only a small portion reflecting positive sentiment. This distribution highlights the prevalent feelings of tension and pessimism among the public when facing disasters like hurricanes. On the other hand, posts on TikTok showed the highest proportion of positive emotions, indicating that although hurricane-related content tends to be negative, TikTok users are more likely to share uplifting or supportive data samples using the platform’s visual and interactive features.

\subsection{Multi-modal Baseline Model Evaluation}\label{baseline}

\begin{table*}[t]
\centering

\setlength{\tabcolsep}{0.5pt} 
\footnotesize
\begin{tabularx}{\textwidth}{c|*{3}{>{\centering\arraybackslash}X>{\centering\arraybackslash}X|}>{\centering\arraybackslash}X>{\centering\arraybackslash}X}
\toprule[1.5pt]
\multirow{2}{*}{\textbf{Category}} & \multicolumn{2}{c|}{\textbf{RoBERTa}} & \multicolumn{2}{c|}{\textbf{BART}} & \multicolumn{2}{c|}{\textbf{ConvBERT}} & \multicolumn{2}{c}{\textbf{ELECTRA}} \\ 
\cmidrule(lr){2-3} \cmidrule(lr){4-5} \cmidrule(lr){6-7} \cmidrule(lr){8-9}
 & Accuracy & Macro F1 & Accuracy & Macro F1 & Accuracy & Macro F1 & Accuracy & Macro F1 \\ 
\midrule
Casualty             & 0.9306 & 0.7439 & 0.9258 & 0.7207 & 0.9303 & 0.7144 & 0.9315 & 0.7431 \\ 
Evacuation             & 0.8009 & 0.6989 & 0.8116 & 0.7110 & 0.8032 & 0.6839 & 0.8040 & 0.6975 \\  
Damage             & 0.7732 & 0.7586 & 0.7664 & 0.7575 & 0.7732 & 0.7585 & 0.7710 & 0.7553 \\ 
Advice           & 0.8344 & 0.6938 & 0.8382 & 0.7115 & 0.8365 & 0.6952 & 0.8326 & 0.6818 \\ 
Request              & 0.8612 & 0.7345 & 0.8599 & 0.7441 & 0.8657 & 0.7463 & 0.8637 & 0.7474 \\ 
Assistance            & 0.7884 & 0.7764 & 0.7888 & 0.7884 & 0.7871 & 0.7767 & 0.7874 & 0.7788 \\ 
Recovery              & 0.8056 & 0.8035 & 0.7980 & 0.7924 & 0.8020 & 0.8001 & 0.8054 & 0.8026 \\ 
\midrule
Linguistic Bias & 0.8395 & 0.7693 & 0.8314 & 0.7552 & 0.8423 & 0.7720 & 0.8417 & 0.7648 \\ 
Political Bias  & 0.9195 & 0.8600 & 0.9216 & 0.8627 & 0.9232 & 0.8653 & 0.9208 & 0.8598 \\ 
Gender Bias     & 0.9631 & 0.5712 & 0.9623 & 0.5776 & 0.9620 & 0.4903 & 0.9620 & 0.4903 \\ 
Hate Speech     & 0.9769 & 0.6898 & 0.9783 & 0.6590 & 0.9785 & 0.6764 & 0.9746 & 0.4936 \\ 
Racial Bias     & 0.9766 & 0.6947 & 0.9767 & 0.6826 & 0.9780 & 0.7085 & 0.9782 & 0.7014 \\ 
\midrule
True/False Information       & 0.8608 & 0.6934 & 0.8628 & 0.6801 & 0.8593 & 0.7156 & 0.8568 & 0.7201 \\ 
\bottomrule[1.5pt]
\end{tabularx}
\caption{Accuracy and Macro F1 for Different Baseline Models}

\label{tab:text-baseline}
\end{table*}

We evaluate the zero-shot performance of three multimodal baseline models on our dataset to establish benchmark performance and assess their capacity to handle disaster-related multimodal posts.
Specifically, we evaluate Phi-4-Multimodal-Instruct~\cite{Phi-4}, LLaVA-Onevision-7B~\cite{LLaVA-OneVision}, and Qwen2.5-VL-7B-Instruct~\cite{qwen2.5-VL} on the annotated labels. These models are selected due to their strong performance on diverse vision-language tasks, as well as their popularity and adoption as standard baselines in prior works~\cite{MiniGPT4-Video2,MiniGPT4-Video1,Phi-4}. 
We conduct binary classification evaluation for every category across the Humanitarian, Bias, and Information Integrity Classes.
We adopt a set of modality-specific preprocessing steps across text, images, and videos to ensure consistent multimodal representation. For the textual content, similar to the work of \citet{shang2024socialdrought}, we remove all URL links and emojis from the content. 
For images, we resize each image to a uniform resolution of 512$\times$512 following standard vision–language preprocessing conventions~\cite{liu2023visual, dai2023instructblip}. For videos, we follow the common practice in video–language research by sampling frames at 1 fps and selecting 8 representative keyframes for downstream analysis~\cite{Howto100m,Videoclip,MiniGPT4-Video1}.
We evaluate the zero-shot performance of all baseline models, meaning that the models are not fine-tuned on our dataset and are instead prompted only with the task definitions. This setup allows us to assess the inherent capability of general-purpose multimodal models to interpret disaster-related posts without any task-specific adaptation.
The experiments are conducted on NVIDIA L40 GPU, and the results are presented in Table~\ref{tab:baseline}. From Tables \ref{tab:DistributionHumanitarianClasses} to \ref{tab:FakeNewsDistribution}, we observe that most of the binary classification tasks exhibit data imbalance, with one class accounting for less than 40\% of the labeled samples. Therefore, we employ two metrics for evaluation: accuracy and macro F1 score, where accuracy focuses on overall performance and the macro F1 score focuses on minority group. 

For most categories, the baseline models achieve accuracy in the 0.6–0.9 range and macro F1 scores in the 0.5–0.7 range, suggesting moderate multimodal understanding but still leaving substantial room for improvement.
One possible reason is that these baseline models are not fine-tuned on disaster-specific content. This limitation is most evident in the true/false information classification task, where the absence of knowledge regarding the 2024 hurricanes leads Phi-4 and Qwen2.5 to perform like random guessing. Moreover, these baseline models are smaller than the Gemini-2.0-Flash that we used for the annotation task, which together contribute to relatively low performance.
Another factor lies in the data processing pipeline where we only adopt 8 keyframes to represent each video due to computational constraints, which leads to the loss of visual features. In addition, the lack of audio information further limits the model's ability to capture important contextual signals, which are often important in video-based content.
These limitations highlight the need for future research on disaster-specific multimodal models that can better leverage multimodal inputs to accurately interpret complex crisis scenarios. In particular, developing smaller but more efficient architectures that can natively integrate text, images, video, and audio will contribute to a more comprehensive and resource-focused understanding of disaster crises.
We additionally report a supervised baseline trained on the textual content of posts in the next section. The results demonstrate strong performance across multiple classification tasks, indicating that the annotations are coherent and that model performance can be improved through task-specific supervision.


\subsection{Text-based Baseline Model Evaluation}\label{text-baseline}

Finally, we evaluate the supervised performance of four text-based baseline models: RoBERTa \cite{roberta}, BART~\cite{bart}, ConvBERT~\cite{ConvBERT}, and ELECTRA~\cite{electra} on the annotated labels. These models are selected due to their strong performance on a wide range of tasks, as well as their popularity and adoption as standard baselines in prior work~\cite{mashkoor2025comparative}. Due to the substantial GPU resources and training time required for supervised multimodal models, we simplify the input to the textual content of social media posts in this evaluation. Similar to the work of \citet{shang2024socialdrought}, we remove all URL links from the content. Additionally, to protect user privacy, all tagged usernames, starting with “@”, are replaced with “@user”. Subsequently, we combine the posts from all four platforms into a single dataset and divide them into 70\% training, 15\% validation, and 15\% testing sets. We train every annotated category using NVIDIA L40S
GPU and the evaluation results are presented in Table~\ref{tab:text-baseline}. We employ two metrics for evaluation: accuracy and macro F1 score, one focuses on the overall performance and the other focuses on the performance of the minority group. From Tables 3 to 5, we observe that most of the binary classification tasks exhibit data imbalance, with one class accounting for less than 40\% of the labeled samples.
To address the class imbalance problem, we use macro F1 score as one of the evaluation metrics for the annotations because it can focus on underrepresented classes.

For most categories, the four baseline models achieve test accuracy over 0.8 and macro F1 score over 0.75, showing strong model performance. However, for some categories, such as gender bias, hate speech, and racial bias, the macro F1 scores are significantly lower. This is mainly due to the small proportion of these categories in the dataset, which is only 2-3\% of the total posts. The small proportion of these categories in the dataset limited the model’s ability to receive sufficient training on these specific labels, thus affecting the macro F1 performance. These findings highlight the challenges of training bias detection models in imbalanced datasets.

\section{Conclusion and Discussion}

This paper introduces a multiplatform and \textbf{M}ultimodal \textbf{A}nnotated Dataset for \textbf{S}ocietal Impact of \textbf{H}urricane (\textbf{MASH}) to facilitate the understanding of societal impact of hurricanes. Specifically, the MASH dataset includes 59,607 relevant social media posts and is multimodal annotated in the three dimensions of Humanitarian Class, Bias Class, and Information Integrity Class. By leveraging annotations in three dimensions, this dataset provides new opportunities for research on the social impact of hurricanes such as the spread of false claims during disaster events, the expression of public sentiment and bias, and the evolution of humanitarian needs over time. We implement systematic analyses to the MASH dataset including correlation analysis, temporal analysis, spatial analysis, baseline model evaluation, and sentiment analysis. Furthermore, we develop an online platform that supports data exploration, provides preliminary analysis, and enables stakeholders to share insights.
The data collection, annotation framework, and online platform introduced in this paper is generalizable to other disaster contexts across diverse social media platforms. 
We will continue to explore responsible and ethical methods to collect rich data on disaster impacts.

\bibliography{CameraReady/LaTeX/aaai2026}
\clearpage
\appendix

\begin{figure*}[t]
\centering
\footnotesize
\begin{tcolorbox}[colback=gray!5, colframe=gray!80!black, title=Prompt]
Read the post, considering text, image, and video together. Determine whether the post is related to an actual hurricane disaster in North America (especially Hurricane Helene and Hurricane Milton) and give your reason. \\ 
Accepted examples:\\
- Hurricane disaster, even if it's not Helene or Milton\\ 
Some typical counter examples:\\
- Miami Hurricanes football team, Carolina Hurricanes hockey team, or other sports names\\
- WWE wrestler called ‘Hurricane’ or other people called 'Hurricane'\\
- Advertisement\\
- Use hurricane as a metaphor to describe other things\\
- The post contains hurricane-related hashtag, but the actual content is not related to the hurricane disaster\\
- Cyclone or Typhoon that does not happen in North America\\ 
Respond with:\\
\{``Judgment'':   "Judgment": True or False,\\
  ``Reason'': ``your detailed explanation based on the post, the examples, and the comparative reasoning between the two viewpoints.''
\}

\end{tcolorbox}
\caption{Prompt for Data Cleaning.}
\label{fig:prompt-clean}
\end{figure*}

\begin{figure*}[t]
\centering
\footnotesize
\begin{tcolorbox}[colback=gray!5, colframe=gray!80!black, title=Prompt]
You are reviewing a social media post related to a hurricane disaster. The post may include text, images, and videos. Carefully examine all available content together.\\ 
\{\textit{Category Definition here}\}
\\ 
Respond in strict JSON format:
\{``Judgment'':   "Judgment": True or False,\\
  ``Reason'': ``Your detailed explanation based on the content of the post.''
\}
\end{tcolorbox}
\caption{Prompt for Humanitarian and Bias Class Annotation.}
\label{fig:prompt-human-bias}
\end{figure*}

\begin{figure*}[t]
\centering
\footnotesize
\begin{tcolorbox}[colback=gray!5, colframe=gray!80!black, title=Prompt]
You are reviewing a social media post related to a hurricane disaster. The post may include text, images, videos, and external links. Carefully examine all available content together.
Your task is to determine whether the post contains false claims, describes truthful information, or is hard to identify due to a lack of verifiable evidence. To make your judgment, search online for reliable sources (e.g., government agencies, major news outlets, or verified fact-checking websites). Use these sources to verify any factual claims made in the post.
\\ 
Label the post as one of the following:\\
- Flase Claims (-1): The post contains or promotes false or misleading information, even if only part of the content is inaccurate.\\
- True Claims (1): The post is accurate and supported by reliable sources.\\
- Hard to Identify (0): The post's truthfulness is difficult to determine (e.g., it describes personal experiences or no trustworthy sources can be found to verify it).
\\ 
Please include the source(s) used to support your reasoning. Provide:\\
- A brief description of the source and how it supports your judgment.\\
- A URL link to the source (e.g., official website, fact-checking report).
\\ 
Respond in strict JSON format:
\{
  ``Judgment'': one of the following integer values: 1 = True Claims, 0 = Hard to Identify, -1 = False Claims\\
  ``Reason'': Your explanation, including supported sources and their descriptions + url links in string format
\}
\end{tcolorbox}
\caption{Prompt for Information Integrity Annotation.}
\label{fig:prompt-info-inte}
\end{figure*}

\begin{figure*}[t]
\centering
\footnotesize
\begin{tcolorbox}[colback=gray!5, colframe=gray!80!black, title=Prompt]
You are reviewing a social media post related to a hurricane disaster. The post may include text, images, and videos. Carefully examine all available content together.\\
\{\textit{Category Definition here}\}
\\
For this specific post, here are two sets of arguments:\\
- Arguments in favor (why the post may be related to a real hurricane disaster): \{true reasons here\}\\
- Arguments against (why the post may NOT be related to a real hurricane disaster): \{false reasons here\}\\
Carefully compare and contrast both sides of the arguments above, and critically assess which side presents a more credible and reasonable explanation based on the content of the post.\\
Consider the relevance, specificity, and plausibility of each argument in the context of the post, and use that analysis to form your final judgment.\\
Respond in strict JSON format:
\{``Judgment'':   "Judgment": True or False,\\
  ``Reason'': ``Your detailed explanation based on the content of the post.''
\}
\end{tcolorbox}
\caption{Prompt for Second Round Annotation Discussion.}
\label{fig:prompt-second}
\end{figure*}
\FloatBarrier   

\begin{table*}[h!]
\centering
\footnotesize
\renewcommand{\arraystretch}{1.2}
\begin{tabular}{c|L{15cm}}
\toprule[1.5pt]
\textbf{Category} & \textbf{Definition} \\
\midrule
\multirow{4}{*}{\textbf{Casualty}} 
&  The post reports people or animals who are killed, injured, or missing during the hurricane. For example:\newline
- Deaths resulting from the hurricane\newline
- Individuals or groups reported as missing \newline
- People or animals who are injured due to storm impact\\
\midrule
\multirow{5}{*}{\textbf{Evacuation}} 
&  The post describes the evacuation, relocation, rescue, or displacement of individuals or animals due to the hurricane. For example:\newline
- People leaving their homes or being urged to evacuate from at-risk areas to safer locations (e.g., temporary shelters, higher ground, or other towns)\newline
- Displacement of communities due to destruction of homes, rising water, or uninhabitable conditions\newline
- Rescue operations carried out by emergency personnel, neighbors, or volunteers\\
\midrule
\multirow{4}{*}{\textbf{Damage}} 
&  The post reports damage to infrastructure or public utilities caused by the hurricane. For example:\newline
- Destruction or severe damage to buildings, homes, or roads\newline
- Power outages, downed power lines, or transformers exploding\newline
- Cell towers, internet lines, or other communication infrastructure being down\\
\midrule
\multirow{4}{*}{\textbf{Advice}} 
&  The post provides advice, guidance, or suggestions related to hurricanes, including how to stay safe, protect property, or prepare for the disaster. For example:\newline
- Safety tips for individuals, families, or pets during a hurricane (e.g., “stay indoors,” “avoid floodwaters”)\newline
- Instructions on how to prepare emergency kits, store food/water, or stock up on supplies\newline
- Advice on how to evacuate safely, choose a shelter, or avoid traffic routes\\
\midrule
\multirow{4}{*}{\textbf{Request}} 
&  The post contains request for help, support, or resources due to the hurricane. For example:\newline
- People asking for rescue or evacuation assistance \newline
- Requests for food, water, medical supplies, or basic necessities\newline
- Posts seeking temporary shelter, housing, or safe relocation options\\

\midrule
\multirow{8}{*}{\textbf{Assistance}} 
& The post contains assistance and support to victims.
This includes both physical aid and emotional or psychological support provided by individuals, communities, or organizations.\newline
Physical and Material Assistance:\newline
- Distribution of relief supplies (e.g., food, water, clothing, medical kits)\newline
- Volunteers helping with debris removal, house repairs, or deliveries\newline
- Government or NGO programs providing direct support (e.g., FEMA aid, Red Cross response)\newline
Emotional or Psychological Support:\newline
- Expressions of solidarity, empathy, or moral support (e.g., “Praying for Florida,” “Stay strong New Orleans”)\newline
- Messages offering comfort, hope, or encouragement to victims and survivors\newline
- Posts acknowledging the suffering of affected communities and standing in support\\
\midrule
\multirow{4}{*}{\textbf{Recovery}} 
& The post describes efforts or activities related to the recovery and rebuilding process after the hurricane. For example:\newline
- Rebuilding homes, businesses, infrastructure, or public spaces after the storm\newline
- Updates on clearing debris, restoring electricity, or repairing roads\newline
- Reopening of schools, government services, stores, or public transport\\
\bottomrule[1.5pt]
\end{tabular}
\caption{Definition for each Humanitarian Category}

\label{tab:def-human}
\end{table*}

\begin{table*}[t]
\centering
\footnotesize
\renewcommand{\arraystretch}{1.2}
\begin{tabular}{c|L{15cm}}
\toprule[1.5pt]
\textbf{Category} & \textbf{Definition} \\
\midrule

\multirow{5}{*}{\textbf{Linguistic Bias}} 
& The post contains biased, inappropriate, or offensive language, with a focus on word choice, tone, or expression. For example:\newline
- Use profanity, vulgarities, or swear words (e.g., “f**k,” “sh*t”) in a way that is excessive, aggressive, or unnecessary\newline
- Show emotionally charged language meant to belittle or ridicule victims, responders, or institutions\newline
- Contain exaggerated, alarmist, or inflammatory wording (e.g., “this hellstorm wiped out everything!” when not supported by evidence)\\
\midrule
\multirow{5}{*}{\textbf{Political Bias}} 
& The post expresses political ideology, showing favor or disapproval toward specific political actors, parties, or policies. For example:\newline
- Promote or criticize political leaders (e.g., mayors, governors, presidents) in relation to the hurricane response or preparedness\newline
- Express support for or opposition to political parties or ideologies\newline
- Frame the hurricane or its response through a politicized lens\\ 
\midrule
\multirow{4}{*}{\textbf{Gender Bias}} 
& The post contains biased, stereotypical, or discriminatory language or viewpoints related to gender. For example:\newline
- Contain sexist language, slurs, or dismissive terms toward certain genders\newline
- Reinforce traditional or harmful gender stereotypes\newline
- Make generalizations about a specific gender \\
\midrule
\multirow{5}{*}{\textbf{Hate Speech}} 
& The post contains language that expresses hatred, hostility, or dehumanization toward a specific group or individual, especially those belonging to minority or marginalized communities. For example:\newline
- Racial or ethnic minorities\newline
- Immigrants or refugees\newline
- Religious minorities\\
\midrule
\multirow{4}{*}{\textbf{Racial Bias}} 
& The post contains biased, discriminatory, or stereotypical statements directed toward one or more racial or ethnic groups. For example:\newline
- Black or African American communities\newline
- Latinx/Hispanic populations\newline
- Asian or Pacific Islander groups\\

\bottomrule[1.5pt]
\end{tabular}
\caption{Definition for each Bias Category}

\label{tab:def-bias}
\end{table*}
\FloatBarrier   

\end{document}